# Non-Contact Breathing Rate Detection Using Optical Flow

Robyn Maxwell, Timothy Hanley, Dara Golden, Adara Andonie, Joseph Lemley, and Ashkan Parsi
*OCTO Sensing Team, Xperi Inc., Galway, Ireland*

**Abstract**

Breathing rate is a vital health metric that is invaluable indicator of the overall health of a person. In recent years, the non-contact measurement of health signals such as breathing rate has been a huge area of development, with a wide range of applications from telemedicine to driver monitoring systems. This paper presents an investigation into a method of non-contact breathing rate detection using a motion detection algorithm, optical flow. Optical flow is used to successfully measure breathing rate by tracking the motion of specific points on the body. In this study, the success of optical flow when using different sets of points is evaluated. Testing shows that both chest and facial movement can be used to determine breathing rate, but to different degrees of success. The chest generates very accurate signals, with a RMSE of 0.63 on the tested videos. Facial points can also generate reliable signals when there is minimal head movement but are much more vulnerable to noise caused by head/body movements. These findings highlight the potential of optical flow as a non-invasive method for breathing rate detection and emphasize the importance of selecting appropriate points to optimize accuracy.

## 1  Introduction

Breathing rate (BR) is a vital physiological signal that can indicate the overall health of a person by providing an insight into their respiratory function and physical fitness, as well as stress and anxiety levels. In part due to the devastation of the 2021 global pandemic, there has been a call for a more comprehensive telemedicine service in recent years. This demand has spiked a huge push for research into the possibility of non-invasive methods of measuring human vital signs such as BR. Contact based measurement of this metric is typically carried out using a wearable sensor like a spirometer or breathing belt however, in many cases, where these methods are not possible, it must be detected through manual counting. The downsides to these methods, although accurate, is that they are intrusive, can lead to the spread of infection and are often not feasible in situations where the person is acutely unwell.

In addition to this, the need for accurate non-invasive BR detection is increasingly present due to recent legislation requiring all EU cars to implement a form of Driver Monitoring System (DMS) from 2026 [M. Bassani, 2023]. The purpose of DMS is to alert drivers to deteriorating health levels, such as drowsiness, stress and distraction and will be instrumental in reducing the number of collisions on the road and improving overall driver safety. For this reason, this study is being carried out by Xperi Inc. to further research in this area.

The majority of existing solutions for this application in terms of optical systems use thermal imaging to detect the BR of a person by analysing the variation in the temperature of the skin caused by respiration. However, the drawbacks of this method are the high cost, lack of spatial resolution and depth perception and the sensitivity to environmental factors. For this reason, this study explores the use of optical flow as a method of detecting BR using both near infrared (NIR) and RGB videos. Optical flow is a computer vision technique that tracks the movement of specific pixels between consecutive frames in a video. The method outlined in this paper compares the accuracy of the detected BR when using specified points on the face compared with points on the chest.

# 2 Breathing Rate Extraction Method

## 2.1 Optical Flow

Optical flow is a computer vision technique that is commonly used to track the motion of objects in a video [Tianqi Guo1 and Allebach1, 2021]. It works under the 'brightness constancy assumption' which is that pixel intensity does not change between consecutive frames. This assumption allows for the displacement of each pixel between frames to be calculated and processed to generate a meaningful signal. In this case, sparse optical flow is used, meaning that motion vectors are calculated only from the specific set of points passed in from the face detector or open pose. The optical flow algorithm used is the Lucas-Kanade feature tracker with a window size of 20 x 20. Testing determined this to be the most effective window size for this study as it could most effectively track pixels of homogeneous texture or colour.

The points to track are detected in the first frame of the video and passed into optical flow. These points are then tracked through each consecutive frame of the video. The difference in position of the y co-ordinate of the points between frames is used to generate the resulting raw signal which is then filtered and used to calculate breathing rate.

This study assesses whether BR can be accurately detected from the face and chest and compares the relative merits of the three methods of detection outlined in further detail below. The first and second method use three points on the face and chest respectively, and the third method uses a triangular grid as points for optical flow to track.

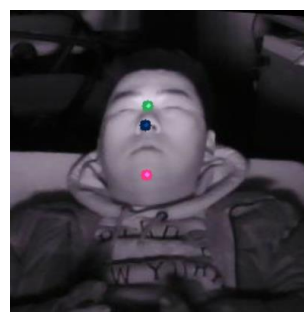

**Figure 1 Points tracked on face**

## 2.2 Face and Chest Point Detection

In order to determine the facial points to track, the KAPAO (Keypoints and Poses as Objects) human pose estimation method was used [McNally et al., 2022]. Face detection is carried out on the first frame of the input video to ascertain the location of the three points of interest. The chosen points are the midpoint between the eyes, the nose, and the chin. These points were identified as those least affected by facial deformation or noise. The points on the lips and eyes were too susceptible to noise and therefore discarded.

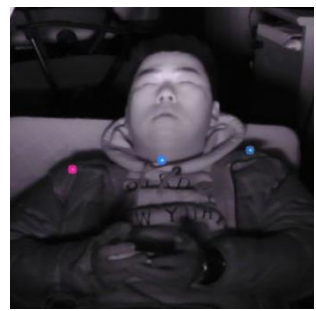

**Figure 2 Points tracked on chest**

The chest points were identified using Open Pose, an open-source computer vision library that uses deep learning to track key points on the human body [Zhe Cao, 2021]. The chest movement was tracked using two similar approaches. The first approach uses the left shoulder, right shoulder, and neck points as arguments for optical flow to track while the second approach uses a triangular grid of points with the shoulder points used the base.

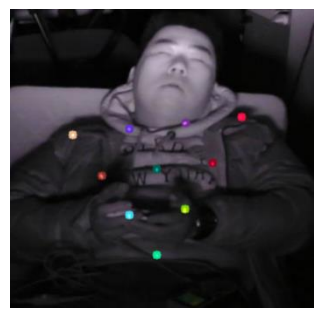

**Figure 3 Triangular chest grid**

## 2.3 Post Processing

To extract the BR signal from the raw output signal of the optical flow, the signal is band passed in the range of 0.1 Hz to 0.5 Hz which corresponds to a BR of 6-30 bpm. The BR of a healthy adult is normally between 10 and 20 bpm. The BR itself is calculated by dividing the number of peaks detected by the peak detector on the filtered signal by the length of the video in seconds and multiplying by 60.

## 2.4 Dataset

This method was tested on two different sets of data in order to capture different poses and rates of respiration. The first set of data consisted of 6 videos, captured internally at Xperi Inc. Each video was 30 seconds in length and consisted of a participant sitting upright in a chair against a neutral background breathing at different rates (14-26 bpm). The subject changes clothes after the first three videos from a patterned shirt to a single colour top to test the effectiveness of optical flow when tracking pixels of the same colour. To compliment this dataset, videos from a publicly available sleep database were incorporated in the testing [Menghan Hu et al., 2018]. This database consists of videos of 12 sleeping participants captured using near infrared (NIR) cameras. From the sleep database, only 6 videos were selected for analysis due to excessive movement by the subject which distorted the results. The purpose of the additional data was to test the success of the method when participants were lying down and to assess whether the method was effective using NIR videos.

## 3 Results

Each method was tested on six videos, three NIR and three from the webcam footage. The videos selected all had range of slight participant movement. This aided in determining the effects of noise on the signal. The resulting graphs were firstly compared against the ground truth signal for accuracy and then against each other to determine signal strength. The chest points achieved an accurate BR signal in each test.

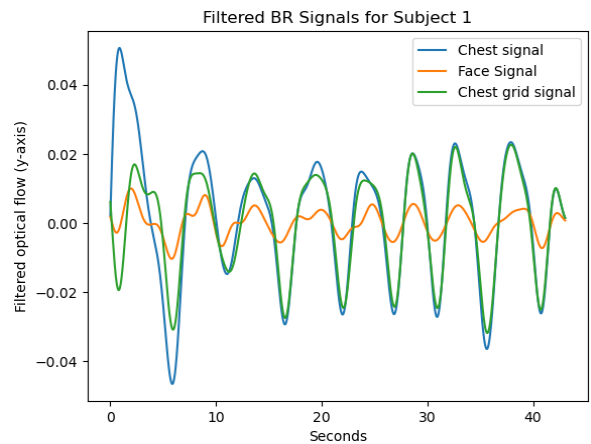

**Figure 4 BR signal for subject 1 (NIR video)**

The resulting signals showed that BR was accurately detectable using each of the three methods, if there was no external noise such as head movement/arm movement. The BR signal from the face provided the weakest signal as overall it was the signal most consistently effected by noise. However it did provide an accurate signal in the videos with the low participant motion and particularly the sleeping videos with the face at a slight angle, as shown in Figure 4.

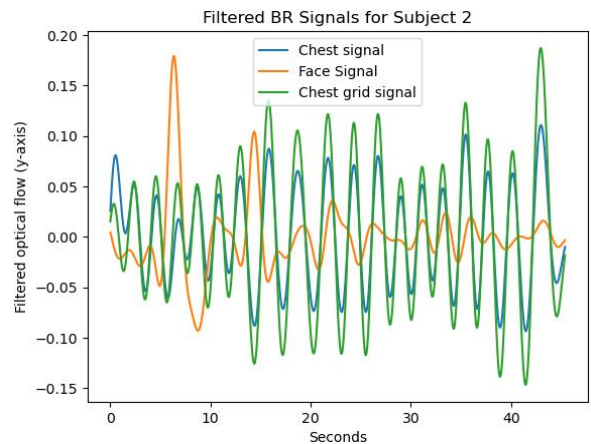

**Figure 5 BR signal for subject 5 (webcam video)**

The signal from the three points on the chest resulted in a strong signal and had the lowest RMSE of the three methods as shown in Table 1 below in all the tests as it could function at a low optical flow window size and was not overly affected by noise. The chest grid provided the strongest signal overall but did not perform as well in the webcam videos where the participant was wearing a solid green jumper. Optical flow had difficulty tracking the points and the window size had to be increased to 40 x 40. This was not an issue with the NIR footage.

|  | Face Points | Chest Points | Chest Grid |
|---|---|---|---|
| **Average RMSE** | 7.0288 | 0.6305 | 0.76511 |

**Table 1 Average RMSE's for each method**

The graph in Figure 5 is an example of a video with a relatively large amount of facial movement. The signal from the face is distorted while the chest signals are both accurate.

Table 1 shows that this method produces a lower RMSE than many other BR detection methods which report RMSE of 1.7 and 2.03 [Jafar Pourbemany and Zhu, 2021] [Tianqi Guo1 and Allebach1, 2021]. The two chest point variations give very low RMSEs with the three-point input being slightly more accurate in detecting the actual BR value. Despite the strength of the results, further development is required to improve the model so that it can function is situations when there is increased movement from the participant. This method performs well on footage with low levels of noise but would not work on a moving participant.

## 4 Conclusion and Further Work

This method of breathing rate detection has produced accurate results and could be further developed to perform accurate real-time non-invasive breathing rate detection. Future work on this will involve improving the robustness of the system against noise caused by participant motion. This could be done by analysing the driver pose and rejecting signals where there a large amount of noise in the frame is evident. It is also possible that a mixture of all three variations could be used in conjunction with each other to create a system that adapts to the motion present in the system and selects the most probable signal with the least noise. In terms of the triangular chest grid, the strength and reliability of the signal could be further improved by preprocessing the image in order to enhance the gradients in the image and provide more distinct points for optical flow to track.